\documentclass{article}%
\usepackage[top=3cm, bottom=3cm, left=3cm, right=3cm]{geometry}
\usepackage{graphicx}
\usepackage{cite}
\usepackage{hyperref}
\usepackage{amsmath}
\usepackage{booktabs}
\usepackage{epstopdf}%
\usepackage{amsfonts}%
\usepackage{amssymb}
\begin{document}

\title{The lightest scalar meson in a simple approach}
\author{T. Wolkanowski and F. Giacosa \\ {\em Institut f\"{u}r Theoretische Physik, Goethe-Universit\"at Frankfurt am Main,} \\ {\em Max-von-Laue-Str. 1, 60438 Frankfurt am Main, Germany}}
\maketitle

\begin{abstract}
We study basic properties of scalar hadronic resonances within a quantum field theoretical toy model. In particular, we focus on the spectral function, the mass and the decay width of the resonance $f_{0}(500)$. In this work, this meson is understood as a seed state in an effective Lagrangian which couples to pions. With such a setup we use the position of the pole on the second Riemann sheet in order to obtain its spectral function. We confirm that $f_{0}(500)$ cannot be described by an ordinary Breit--Wigner function, and that a more complicated structure is needed.

\end{abstract}

\section{Introduction}
The issue of scalar mesons has been the subject of vivid debate among the physical community for a long time because their identification and explanation in terms of quarks and gluons is difficult, see e.g. Refs. \cite{amslerrev,scalars,denis,dick} and refs. therein. Especially some of those particles possess large decay widths, several decay channels and a huge background \cite{beringer}. During the past decades a large number of theoretical approaches were invented to handle these problems and to extract particle information from experimental data, in particular with the aim of understanding the lightest scalar state, the resonance $f_{0}(500)$. Note, this resonance was identified with the scalar $\sigma$-field of first $\sigma$-models introduced by Gell-Mann and L\'{e}vy \cite{levy}, in which they referred to a field corresponding to a spinless meson that was introduced by Schwinger \cite{schwinger}. Nowadays, modern versions of the $\sigma$-model \cite{denis,dick} identify the $\sigma$-field (that is, the chiral partner of the pion) with the scalar resonance $f_{0}(1370)$. Conversely, the state $f_{0}(500)$ is now often interpreted as a molecular state or a so-called tetraquark state \cite{maiani,thooft,giacosaTetra}, a hypothetical mesonic structure first suggested by Jaffe \cite{jaffe}.

New theoretical efforts were made by T\"{o}rnqvist and Roos \cite{roos} by locating the pole position of $f_{0}(500)$ when fitting the whole light scalar nonet to scattering data. It is also possible to determine the $\sigma$-pole by using Roy equations with crossing symmetry, analyticity and unitarity -- many works going in this direction were published in the last years and each has found comparable results. For instance, Caprini \emph{et al.} have stated in Ref. \cite{caprini} that the $\pi\pi$-scattering amplitude contains a pole with the quantum numbers of $f_{0}(500)$ and calculated its mass and decay width within small uncertainties: $\sqrt{s_{\text{pole}}}=(441_{-8}^{+16}-i272_{-12.5}^{+9})$ MeV. Dispersive analysis of Garc\'{i}a-Mart\'{i}n \emph{et al.} in Ref. \cite{kaminski} shows a similar result: $\sqrt{s_{\text{pole}}}=(457_{-13}^{+14}-i279_{-7}^{+11})$ MeV. Both specifications should make clear that the $f_{0}(500)$ state is a very broad resonance with a decay width $\Gamma\sim M$, hence it cannot be parameterized by ordinary methods, for example by using a simple Breit--Wigner distribution function.

In this paper, which is based on the findings of Ref. \cite{giacosa38}, we concentrate on a simple quantum field theoretical toy model: a scalar state can decay into two (pseudo-)scalar ones (alias the pions). All but one of the free parameters inside the model are fixed by applying the resonance pole from either of the two references mentioned above as the propagator pole on the second Riemann sheet. Since the position of the pole is influenced by hadronic loop contributions of the two (pseudo-)scalar particles, a function of the cutoff scale $\Lambda$, we vary this parameter and obtain the spectral functions.

\section{The model}
Our model consists of two scalar fields, $S$ and $\phi$, described by a Lagrangian of the form
\begin{equation}
\mathcal{L}=\frac{1}{2}(\partial_{\mu}S)^{2}+\frac{1}{2}(\partial_{\mu}\phi)^{2}-\frac{1}{2}M_{0}^{2}S^{2}-\frac{1}{2}m^{2}\phi^{2}+gS\phi^{2} \ . \label{equation_lagrangian}
\end{equation}
The same model was already studied by Veltman \cite{veltman} as well as by Giacosa and Pagliara \cite{giacosaSpectral} regarding its spectral function: it contains an interaction term for a one-channel decay process $S\rightarrow\phi\phi$. In the one-loop approximation, the hadronic loop contributions from the $\phi$-fields appear in the inverse expression of the full interacting propagator of the scalar field $S$ after Dyson resummation,
\begin{equation}
\Delta_{S}^{-1}(p^{2})=p^{2}-M_{0}^{2}+(\sqrt{2}g)^{2}\Sigma(p^{2}) \ , \label{equation_momentumfullprop}
\end{equation}
where the emerging self-energy function $\Sigma(p^{2})$ represents the two $\phi$-particles inside a mesonic loop (incoming and outgoing momentum is denoted by $p$):
\begin{equation}
\Sigma(p^{2})=-i\int\frac{d^{4}q}{(2\pi)^{4}}\frac{f_{\Lambda}^{2}(q)}{\left(\frac{p}{2}+q\right)  ^{2}-m^{2}+i\epsilon}\frac{1}{\left(  \frac{p}{2}-q\right)  ^{2}-m^{2}+i\epsilon} \ . \label{equation_hadronicloop}
\end{equation}
The regularization function $f_{\Lambda}(q)$, which depends on a UV cutoff scale $\Lambda$, was introduced to make the otherwise logarithmic divergent integral finite (it can be included at the Lagrangian level by rendering it nonlocal, e.g. Ref. \cite{nonlocal}). Since we deal with an effective Lagrangian in the low-energy regime to study light mesons, the mass scale of our model is determined: it is reasonable to set $\Lambda$ between 1 and 2 GeV. The integral (\ref{equation_hadronicloop}) is then evaluated by assuming the regularization function to depend only on the magnitude of the three-momentum; we adopt the choice $f_{\Lambda}(q)\equiv f_{\Lambda}(\mathbf{q}^{2})=(1+\mathbf{q}^{2}/\Lambda^{2})^{-1}$, yielding an exponentially decreasing interaction strength between the two $\phi$-particles for increasing distance.

According to the K\"{a}ll\'{e}n--Lehmann spectral representation, the spectral function $d_{S}$ is known to be the negative imaginary part of the propagator taken slightly above the real axis of the complex $p^{2}$-plane. In our case it is more practical to consider the propagator in the complex $x$-plane, where $x=\sqrt{p^{2}}$ is a running mass variable:
\begin{equation}
d_{S}(x=\sqrt{p^{2}})=-\frac{2x}{\pi}\operatorname{Im}\Delta_{S}(x^{2}+i\varepsilon) \ .
\end{equation}

We now continue the propagator into the second Riemann sheet by exploiting the idea of a Riemann surface and insert each one of the above poles to determine the coupling constant $g$ and the bare mass parameter $M_{0}$. The latter corresponds to the mass of the free seed state as obtained from the Lagrangian (\ref{equation_lagrangian}) in the case of $g=0$.

When the coupling constant is large, the mass and width of the resonance can be successfully determined by the position of the complex pole of the full interacting propagator \cite{peierls,hoehler,levypoles,aramaki,landshoff}. Indeed, large values of the coupling constant to intermediate hadronic states is one of the reasons why the scalar sector is not easily described by the naive quark model \cite{pennington}.

\section{Results and discussions}
The spectral function is determined for the poles of Ref. \cite{caprini} and \cite{kaminski}, and for two different values of the cutoff parameter: $\Lambda=1$ GeV and $\Lambda=2$ GeV. The mass of the $\phi$-fields (alias the pions) is set to $m=m_{\pi}=135$ MeV. We list the outcome of the numerical analysis in Table \ref{table_parameters}.

For a fixed cutoff we obtain very similar values for $M_{0}$ and $g$ for both poles: this is expected, since the two poles differ only marginally. When comparing the values between two different cutoffs, we note that, while the coupling constant hardly changes, the bare mass increases sizably for a larger cutoff. From a phenomenological point of view it is natural to think of $M_{0}$ as being highly influenced by the strong coupling to intermediate hadronic states and the creation of a mesonic cloud, respectively. Such kind of mechanism usually causes the physical mass (as the real part of the propagator pole) to be smaller than $M_{0}$, especially when increasing the coupling. A greater value for the cutoff enhances this influence.
\begin{table}[t]
\center
\begin{tabular}
[c]{lcccc}
\toprule & \multicolumn{2}{c}{$\Lambda=1$ GeV} & \multicolumn{2}{c}{$\Lambda=2$ GeV}\\
\midrule Used pole & $M_{0}$ [GeV] & $g$ [GeV] & $M_{0}$ [GeV] & $g$ [GeV]\\
\midrule Caprini \emph{et al.} \cite{caprini} & 0.468 & 2.013 & 0.561 & 2.034\\
Garc\'{i}a-Mart\'{i}n \emph{et al.} \cite{kaminski} & 0.478 & 2.086 & 0.576 & 2.103\\
&  &  &  &
\end{tabular}
\caption{Numerical results for the bare mass parameter $M_{0}$ and the coupling constant $g$ in dependence of the cutoff $\Lambda$.}
\label{table_parameters}
\end{table}

\begin{figure}[tbh]
\centerline{\includegraphics[width=11.0cm]{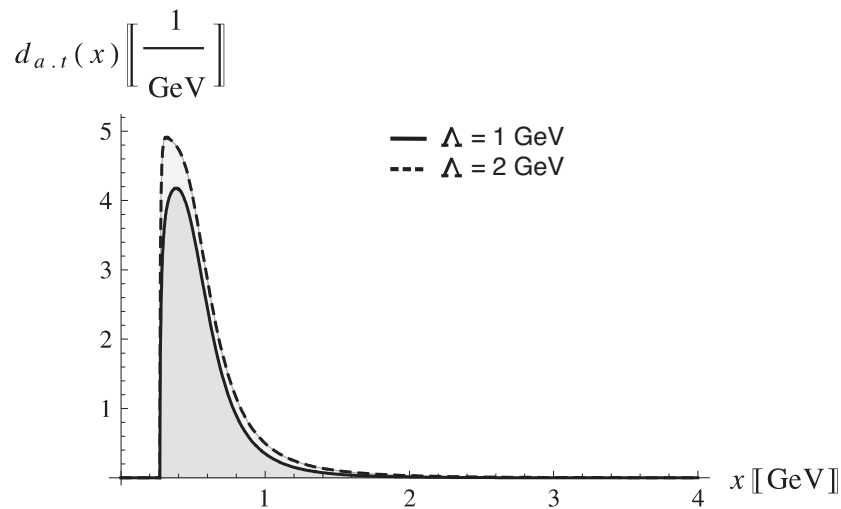}}\caption{Continuous part of the spectral function for the case $\Lambda=1$ GeV (solid line) and $\Lambda=2$ GeV (dashed line) for the pole of Garc\'{i}a-Mart\'{i}n \emph{et al.}}
\label{figure_spectral}
\end{figure}

We show in Fig. \ref{figure_spectral} the spectral function above threshold, denoted as $d_{a.t}$, for the resonance pole of Garc\'{i}a-Mart\'{i}n \emph{et al.} and the parameters obtained above (the higher cutoff corresponds to the dashed curve). In both cases the typical Breit--Wigner shape for an ordinary resonance is clearly absent: the peak is strongly shifted to the left and the whole structure is asymmetric. Besides that, the spectral function is not normalized to one because a simple pole emerges on the first Riemann sheet below threshold, see also Ref. \cite{giacosa38} for further details about this pole and its possible, albeit speculative, physical meaning. Increasing the cutoff reduces the lack of normalization. In summary, a detailed study of the cutoff dependence in our model is desirable and a possible outlook for future work.

In conclusion, we have studied the spectral function of the lightest scalar meson $f_{0}(500)$ within a simple quantum field theoretical model. In the future, one should also include the constrains of chiral symmetry and the rich hadron phenomenology by using the model developed in Refs. \cite{denis,dick}.

\bigskip

\textbf{Acknowledgments:} The authors thank G. Pagliara, D. H. Rischke and J. R. Pel\'{a}ez for useful discussions.

\end{document}